\documentclass[reprint,secnumarabic,amssymb, nobibnotes, aps, prb]{revtex4-1}

\usepackage{graphicx}
\usepackage{graphics}
\usepackage[percent]{overpic}
\graphicspath{ {images/} }
\usepackage{amsmath}
\usepackage{mathtools}

\usepackage[utf8]{inputenc}
\usepackage{amsfonts}
\usepackage{amssymb}
\usepackage{float}
\usepackage{comment}
\usepackage{color,soul}
\usepackage{textcomp}
\usepackage{subcaption}
\usepackage{hyperref}
\usepackage[dvipsnames]{xcolor}
\usepackage{physics}
\newcommand{\vQ}{{\bf Q}}
\newcommand{\vk}{{\bf k}}

\renewcommand{\vr}{{\bf r}}

\begin{document}

%%%%%%%%%%%%%%%%%%%%%%%%%%%%%%%%%%%%%%%%%%%%%%%%%%%%%%%%%%%%%%%%%%%%%%%%%%%%%%%%%%%%%%%%%%%%%%%%%%%%%%%%
\title{Anisotropic Thermal Transport in Superconductors with Coexisting Spin Density Waves}%

%\author{Sean Peterson, Sourav Sen Choudhury, Yves Idzerda}%
\author{Sean F Peterson}
\author{Sourav Sen Choudhury}
\author{Yves Idzerda}
%\email[REVTeX Support: ]{revtex@aps.org}
\affiliation{Department of Physics, Montana State University, Bozeman, Montana 59717, USA}
\date{\today}
\begin{abstract}
        \noindent{Thermal conductivity measurements can provide key and experimentally verifiable insight into the electronic transport of unconventional superconductors. In this work, electronic thermal transport of two-dimensional tight-binding metallic systems with coexisting $d$-wave superconducting (SC) and antiferromagnetic spin density wave (SDW) orders with nesting vector $\mathbf{Q} = (\pi/2,\pi/2)$ or $(\pi,0)$ are considered. The coexisting SC and SDW orders are modelled at the mean-field level. Thermal conductivities are numerically calculated within Boltzmann kinetic theory in the weak impurity scattering (Born) limit. These SDW nesting vectors are chosen for their unique property of reconstructing the Fermi surface (FS) parallel to $\mathbf{Q}$ and preserving the metallic FS perpendicular to $\mathbf{Q}$. This leads to anisotropic electronic thermal conductivities parallel and perpendicular to $\mathbf{Q}$, which also depend on the presence or absence of additional gapless excitations exclusive to the coexistence phase. It was found that the $\mathbf{Q} = (\pi/2,\pi/2)$ and $(\pi,0)$ SDW systems exhibit equivalent electron transport relative to $\mathbf{Q}$. These systems also had equivalent electron transport when coexisting with a $d$-wave SC gap when $\Delta_{\mathbf{k}}$ had the same symmetry class under translations of $\mathbf{Q}$.}
\end{abstract}
\maketitle
%\tableofcontents

%%%%%%%%%%%%%%%%%%%%%%%%%%%%%%%%%%%%%%%%%%%%%%%%%%%%%%%%%%%%%%%%%%%%%%%%%%%%%%%%%%%%%%%%%%%%%%%%%%%%%%%%
\section{Introduction}

Among the most studied candidates for high temperature superconductors (SCs) are the cuprates\cite{taillefer}, iron pnictides\cite{wen}, and iron chalcogenides\cite{hsu}. A common feature for these families of materials is that they have quasi-two-dimensional sheets of transition metal atoms (either Cu or Fe) in a square-lattice resulting in cylindrical Fermi surfaces\cite{lebegue}$^,$\cite{singh} (FSs) that can be treated as two-dimensional systems (since they are largely $k_z$-independent). Due to the layered structure of these quasi-two-dimensional sheets, it is possible to grow single superconducting layers on a substrate and study superconductivity strictly in two-dimensions\cite{yu}$^,$\cite{ge}$^,$\cite{bollinger}. It is important to note that many of these high-$T_c$ SCs are unconventional in nature.

Unconventional superconductors often have phase diagrams with multiple broken symmetry phases which depend on material properties such as electron or hole doping concentration\cite{basov}$^,$\cite{sachdev}. One of the more common broken symmetry states that superconductivity can coexist with is an antiferromagnetic (AF) state which couples quasiparticle states in different parts of the Brillouin Zone by a nesting vector, $\vQ$, forming a spin density wave (SDW) state\cite{lake}$^,$\cite{demler}$^,$\cite{harrison}$^,$\cite{tan}. While in this work only the interplay between SDW and SC orders will be investigated, the SDW state will often be preceded by a structural transition from a tetragonal to an orthorhombic\cite{mcguire} or a monoclinic\cite{de_la_cruz} lattice thus breaking the fourfold rotational symmetry ($C(4)$) of the crystal. This structural transition can result in an Ising nematic phase\cite{nie} and its effects on electronic thermal transport in SC systems has been previously discussed\cite{senchoudhury2}. Including effects of a structural transition with SDW and SC ordering is beyond the scope of this work. While it is assumed here that the $C(4)$ symmetry of the underlying structural square lattice is preserved, the magnetic structure imposed on the lattice by the existence of striped AF ordering reduces the $C(4)$ symmetry of the unit cell\cite{taillefer} to that of a twofold rotational symmetry ($C(2)$) in the magnetic cell. This broken symmetry is often reflected in the transport properties of such materials\cite{ran}. 

To better understand these unconventional SCs, thermal conductivity measurements are an invaluable tool for probing the transport properties of materials\cite{shakeripour}$^,$\cite{hashimoto}. In normal metals, the electronic thermal conductivity at low temperatures is dominated by electron scattering off impurities, and results in a linear temperature dependence which is well understood within the framework of semiclassical transport theory based on the Boltzmann kinetic equation\cite{ziman}. In conventional SCs the entire FS is gapped and the thermal conductivity is known to decrease exponentially\cite{bardeen} as $T \rightarrow 0$. However in unconventional SCs (such as $d$-wave SCs) the thermal conductivity is known to have a linear $T$-dependence at low-$T$ in the limit of weak impurity scattering, similar to a normal metal due to the existence of zero-energy quasiparticle excitations (nodes) on the FS\cite{arfi}$^,$\cite{graf}. The band topology in the vicinity of these nodes is of utmost importance as it determines the quasiparticle velocities, which can drastically change the transport properties of a material. For example, it has been shown\cite{senchoudhury} that two types of $d$-wave SC ($d_{xy}$ vs. $d_{x^2-y^2}$) have very different thermal conductivities on tight-binding FSs due to different Fermi velocities and local densities of states at the nodes. Electron transport within SDW materials was observed\cite{kim} to follow suppressed Fermi liquid behavior, and as such has an electronic thermal conductivity that is linear in $T$, but diminished from the normal metallic state thermal conductivity\cite{steckel}$^,$\cite{chatterjee}. Thermal transport in $d$-wave superconducting materials with density waves, such as charge density waves or spin density waves, which reduce the $C(4)$ rotational symmetry to a $C(2)$ rotational symmetry have been shown to exhibit anisotropic thermal transport at low-$T$\cite{durst}$^,$\cite{schiff}.

In the cuprates, the superconducting gap is known to have $d$-wave symmetry\cite{damascelli}. Superconductivity can be preceded by a commensurate SDW order with nesting vector $\vQ = (\pi,\pi)$\cite{lake}$^,$\cite{demler}$^,$\cite{tranquada} (also known as the AF1 state\cite{fang}). This reconstructs\cite{laliberte}$^,$\cite{helm} the metallic FS with quasiparticle pockets located at the $M$ points in the BZ\cite{moon} while preserving the $d$-wave SC symmetry nodes which are the main contributors to the transport properties in the clean limit\cite{chatterjee}. The $d$-wave SC state was also found to coexist with the $\vQ = (\pi,0)$ SDW state in the underdoped region of two-dimensional Hubbard model\cite{zheng}, which is often used to model the cuprates. In thin film cuprates\cite{yu}$^,$\cite{bollinger} the behavior of the bulk phase was preserved in monolayers, including the high-$T_c$ value at optimal doping, indicating that SC in the cuprates is inherently a two-dimensional phenomena. Some cuprate materials have been measured to exhibit anisotropic in-plane electronic thermal conductivities, where electrons preferentially travel along one crystallogrphic direction over another\cite{cohn}$^,$\cite{sun} due to electronic inhomogeneities. Additionally, quasi-one-dimensional electronic thermal transport mediated by spin fluctuations was also observed in the cuprates\cite{sologubenko}.

In the iron pnictides, an unconventional superconducting gap may emerge out of a commensurate SDW state with nesting vector $\vQ = (\pi,0)$\cite{de_la_cruz}$^,$\cite{zhao3}$^,$\cite{chen}$^,$\cite{kimber}$^,$\cite{huang} (also known as the AF2 state\cite{fang}). This results in a striped AF which reduces the $C(4)$ symmetry of the crystal lattice to a magnetic cell with $C(2)$ symmetry\cite{zhao}. In such materials it has been shown that the DC electric conductivity within the Drude model is highly anisotropic between the conductivity parallel and perpendicular to $\vQ$ (i.e. $\sigma_{xx} \neq \sigma_{yy}$)\cite{ran}. 

While SC often arises out of an AF state in the cuprates and iron pnictides, iron chalcogenides lack AF ordering\cite{mizuguchi} in bulk. However, FeSe monolayers can exhibit SDW ordering when grown on substrates that increase the spacing between Fe atoms due to epitaxial strain\cite{tan}. These strained FeSe monolayers have been measured\cite{ge} to have greatly enhanced transition temperatures ($T_c$), when compared to those measured in bulk\cite{hsu}. High-$T_c$ superconductivity in these strained FeSe monolayers is likely due to the presence of SDWs in the material enhancing the SC state\cite{tan}. In some iron chalcogenides, the SDW nesting vector was found to be the commensurate nesting vector $\vQ = (\pi/2,\pi/2)$\cite{qiu}$^,$\cite{li} (also known as the AF3 state\cite{fang}).

In this work, the single band electronic transport properties of two cases are considered. The first being a collinear commensurate SDW state with nesting vector $\vQ = (\pi/2,\pi/2)$ (AF3) coexisting with the $d$-wave singlet pairing SC states: $d_{x^2-y^2}$ and $d_{xy}$. The second case is a similar collinear commensurate SDW state, but with a nesting vector of $\vQ = (\pi,0)$ (AF2) coexisting with the same $d$-wave singlet SC pairing states. The effects arising from the multiplicity of bands were not considered and therefore orbital degrees of freedom were neglected. Thus the present analysis is not directly applicable to iron-based SCs (where orbital mixing plays a significant role), but is more relevant for cuprate SCs which can be accurately modeled with a single band model\cite{foley}$^,$\cite{xu}. Nevertheless, the analysis given below provides important insights regarding the interplay of SC and SDW orders and their impact on thermal transport properties of such systems, particularly in the coexistence phase.

%%%%%%%%%%%%%%%%%%%%%%%%%%%%%%%%%%%%%%%%%%%%%%%%%%%%%%%%%%%%%%%%%%%%%%%%%%%%%%%%%%%%%%%%%%%
\section{Model and Formalism}

\subsection{\label{sec:H} Hamiltonian}

In this work the normal state metallic tight-binding Hamiltonian will be considered:

\begin{equation}
    H_0 = \sum_{\vk,\sigma} \xi_{\vk} \hat{a}^\dag_{\vk \sigma} \hat{a}_{\vk \sigma}
\end{equation}

\begin{figure}[b!]
\begin{centering}
\includegraphics[width=8.5cm]{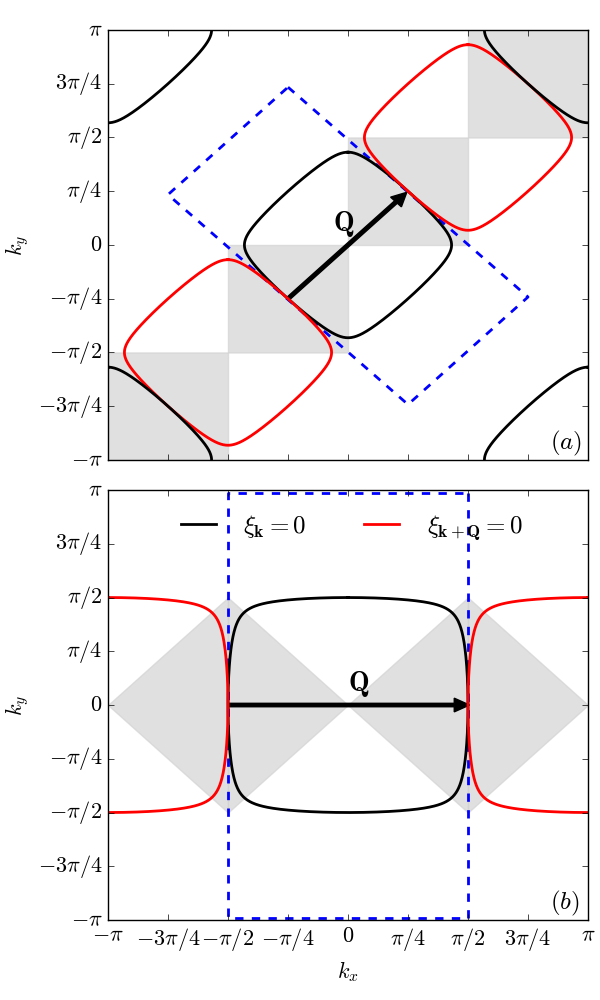}
\caption{Fermi surface nesting between tight-binding FSs (black curves) and the FSs translated by $\vQ$ (red curves) for SDW orders with the nesting vector: (a) $\vQ = (\pi/2,\pi/2)$ (b) and $\vQ = (\pi,0)$ where the FS reconstruction due to SDW ordering occurs in the gray regions and the normal state FS is preserved in the white regions. These SDW states also result in magnetic cells which are larger than the unit cells of the underlying lattice, this results in the periodicity in $\vk$-space being reduced from the square FBZ to the reduced Brillouin Zone seen as the blue dashed curves.} 
\label{fig:SDW_nesting}
\end{centering}
\end{figure}

\noindent with the 2D inversion-symmetric ($\xi_{\vk} = \xi_{-\vk}$) dispersion relations 

\begin{equation}
\begin{split}
    \xi^{(1)}_{\vk} &= \mu - t_1 (\cos2k_x + \cos2k_y) - t_2 \cos2k_x \cos2k_y
\\
    \xi^{(2)}_{\vk} &= \mu - t_1(\cos(k_x - k_y) + \cos(k_x + k_y)) 
\\    
    & - t_2\cos(k_x - k_y)\cos(k_x + k_y)
\end{split}
\end{equation}

\noindent where $\mu$ is the chemical potential, $t_1$ is the nearest neighbor hopping, and $t_2$ is the next-nearest neighbor hopping on a two-dimensional square lattice of spacing one (a = 1), all of which are in units of the N\'eel temperature ($T_N$). The chemical potential was set to zero ($\mu = 0$) and the hopping parameters were set to $t_1 = 100 T_N$ and $t_2 = 10 T_N$ for both dispersion relations, consistent with previous calculations\cite{kato}$^,$\cite{anderson} and experiments\cite{meevasana}$^,$\cite{kordyuk} in literature. For $\xi^{(1)}_{\vk}$ this results in the weak metallic FS ($\xi_\vk^{(1)} = 0$) with quasiparticle pockets centered at the $\Gamma$ and $M$ points in the first Brillouin Zone (FBZ) which can be seen as the black curves in FIG. ~\ref{fig:SDW_nesting}(a). For $\xi^{(2)}_{\vk}$, this results in a metallic FS ($\xi_\vk^{(2)} = 0$) with a quasiparticle pocket centered around the $\Gamma$ point in the FBZ which can be seen as the black curve in FIG. ~\ref{fig:SDW_nesting}(b). The first dispersion relation, $\xi_\vk^{(1)}$, represents a system where the FS is translated by the SDW nesting vector $\vQ = (\pi/2,\pi/2)$ and overlaps with the original FS at the edge of the reduced Brillouin Zone (RBZ) as can be seen in FIG. ~\ref{fig:SDW_nesting}(a). The second dispersion relation, $\xi_\vk^{(2)}$, represents a SDW nesting vector $\vQ = (\pi,0)$ as can be seen in FIG. ~\ref{fig:SDW_nesting}(b). 

\begin{figure}[b!]
\begin{centering}
\includegraphics[width=8.5cm]{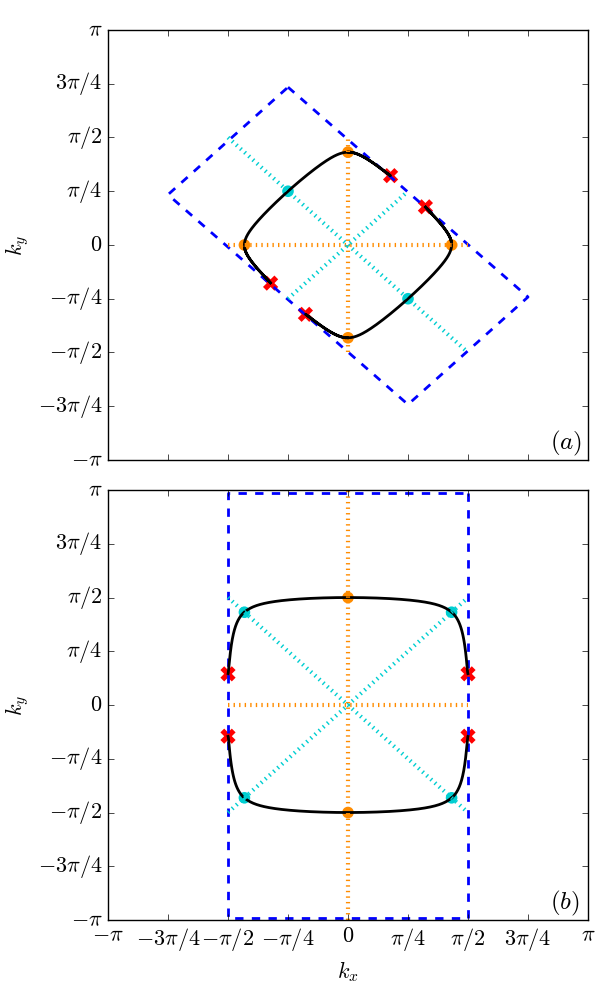}
\caption{FSs reconstructed by SDW ordering (solid black curves) when $M = 2T_N$ for the nesting vectors (a) $\vQ = (\pi/2,\pi/2)$ and (b) $\vQ = (\pi,0)$. The $d$-wave nodal lines are represented with dotted cyan and orange lines for $d_{x^2-y^2}$ and $d_{xy}$ respectively. Cyan and orange points represent the locations of the $d_{x^2-y^2}$ and $d_{xy}$ nodes respectively when their nodal lines cross the reconstructed FS. Red X's show the locations of additional mixing nodes that occur only when the SC gap is even under translations of $\vQ$ ($\Delta_{\vk+\vQ} = \Delta_\vk$).} 
\label{fig:SC_nodes}
\end{centering}
\end{figure}

The emergence of SDW ordering in these metallic systems occurs below the N\'eel temperature and is the result of a striped antiferromagnetic (AF) system. This broken symmetry is also reflected in the shape of the RBZ which can be seen as the blue dashed lines in FIG. ~\ref{fig:SDW_nesting}. Due to this reduced rotational symmetry, the $\xi_\vk$ and $\xi_{\vk+\vQ}$ FSs only overlap in the direction parallel to the nesting vector $\vQ$, leading to a FS reconstruction parallel to $\vQ$ while preserving the normal state tight-binding FS perpendicular to $\vQ$. This reconstructed FS can be seen by the black curve in FIG. ~\ref{fig:SC_nodes}. 

To begin studying thermal transport in a system with coexisting SDW and SC orders, the Hamiltonian is modelled at the mean-field level\cite{kato}:

\begin{equation}
    \begin{split}
    H &= H_0 + H_{\text{SDW}} + H_{\text{SC}} \\
    H_{\text{SDW}} &= \frac{1}{2} \sum_{\vk,\sigma} \sigma M \big(\hat{a}_{\vk,\sigma}^\dagger \hat{a}_{\vk+\vQ,\sigma} + \text{H.c.} \big) \\
    H_{\text{SC}} &= \frac{1}{2} \sum_{\vk,\sigma} \sigma \Delta \mathcal{Y}_\vk \big(\hat{a}_{\vk,\sigma}^\dagger \hat{a}_{-\vk,-\sigma}^\dagger + \text{H.c.} \big)
    \end{split}
\end{equation}
\noindent where the mean-field order parameters are defined by:
\begin{equation}
\begin{split}
    M &= -\frac{V_{\text{SDW}}}{2} \sum_{\vk,\sigma} \sigma \langle \hat{a}_{\vk+\vQ,\sigma}^\dagger \hat{a}_{\vk,\sigma} \rangle \\ 
    \Delta &= -V_{\text{SC}} \sum_{\vk} \mathcal{Y}_\vk \langle \hat{a}_{-\vk,\downarrow}^\dagger \hat{a}_{\vk,\uparrow}^\dagger \rangle
\end{split}
\end{equation}

\noindent These order parameters were found self-consistently by a method outlined in Appendix ~\ref{sec:OPselfConsistency} and superconductivity was assumed to arise out of the SDW ordering ($T_N > T_c$), consistent with phase diagrams for iron-based\cite{wang}$^,$\cite{chu} and cuprate\cite{haug} superconductors. The results of these self-consistency calculations for the order parameters can be seen in FIG. ~\ref{fig:deltaM}. Here a collinear sinusoidal SDW system was considered with a spatial magnetization of $\mathbf{m}(\vr) = 2M \hat{z} \cos{\vQ\cdot\vr}$. SDW ordering couples electron states of parallel spins whose momenta differ by the nesting vector $\vQ$. For the SC order, only singlet electron pairing is considered and $\Delta_\vk = \Delta \mathcal{Y}_\vk$, where $\mathcal{Y}_{\vk}$ is a basis function compatible with the square symmetry inherent to the lattice. The basis functions considered were: $d_{x^2-y^2}$ ($\mathcal{Y}_\vk \propto \cos k_x - \cos k_y$) and $d_{xy}$ ($\mathcal{Y}_\vk \propto \sin k_x \sin k_y$) and qualitative illustrations of these SC gap structures can be seen in FIG. ~\ref{fig:SC_symm} on the two normal state FSs.

The mean-field Hamiltonian for $T < T_c$ and $\Delta \neq 0$ can be written in matrix form in the Nambu basis as:

\begin{equation*}
    H^{(\sigma)} = \frac{1}{2} \sum_{\vk}\big(\hat{\mathbf{\Psi}}_{\vk}^{\text{n}}\big)^{\dag} \hat{\mathcal{H}}^{(\sigma)}_{\vk} \hat{\mathbf{\Psi}}_{\vk}^{\text{n}}
\end{equation*}

\noindent the spin-dependent Hamiltonian matrix in the regions of the RBZ where the FS reconstructs due to SDW ordering (see gray shaded regions in FIG. ~\ref{fig:SDW_nesting}) can be written as:

\begin{equation}
    \hat{\mathcal{H}}_{\vk}^{(\sigma)} = 
    \begin{pmatrix}
    \xi_{\vk} & \sigma\Delta_{\vk} & \sigma M & 0 \\
    \sigma\Delta_{\vk} & -\xi_{\vk} & 0 & \sigma M \\
    \sigma M & 0 & \xi_{\vk+\vQ} & \sigma\Delta_{\vk+\vQ} \\
    0 & \sigma M & \sigma\Delta_{\vk+\vQ} & -\xi_{\vk+\vQ} \\
    \end{pmatrix}
\end{equation}

\noindent and reduces to a pure SC Hamiltonian in regions where the normal state FS is preserved (see unshaded regions in FIG. ~\ref{fig:SDW_nesting}):

\begin{equation}
    \hat{\mathcal{H}}_{\vk}^{(\sigma)} = 
    \begin{pmatrix}
    \xi_{\vk} & \sigma\Delta_{\vk} & 0 & 0 \\
    \sigma\Delta_{\vk} & -\xi_{\vk} & 0 & 0 \\
    0 & 0 & \xi_{\vk+\vQ} & \sigma\Delta_{\vk+\vQ} \\
    0 & 0 & \sigma\Delta_{\vk+\vQ} & -\xi_{\vk+\vQ} \\
    \end{pmatrix}
\end{equation}

\noindent where \big($\hat{\mathbf{\Psi}}_{\vk}^{\text{n}}\big)^{\dag} = (\hat{a}_{\vk,\sigma}^{\dag},\hat{a}_{-\vk, -\sigma},\hat{a}_{\vk+\vQ,\sigma}^{\dag},\hat{a}_{-\vk-\vQ, -\sigma})$ is the Nambu vector for the normal state. The eigenvalues of $\hat{\mathcal{H}}_{\vk}$ in the shaded regions when SC and SDW orders coexist are the quasiparticle energies $\pm E^{(1,2)}_{\vk}$

\begin{equation}
    E^{(1)}_{\vk} = \sqrt{\Gamma_{\vk} + 2\Lambda_{\vk}}, \hspace{.25in} E^{(2)}_{\vk} = \sqrt{\Gamma_{\vk} - 2\Lambda_{\vk}}
    \label{eq:coexistEigs}
\end{equation}

\begin{align*}
    \Gamma_{\vk} &= (\xi_{\vk}^{+})^2 + (\xi_{\vk}^{-})^2 + (\Delta_{\vk}^+)^2 + (\Delta_{\vk}^-)^2 + M^2 \\ 
    \Lambda_{\vk} &= \sqrt{(\xi_{\vk}^+ \xi_{\vk}^- + \Delta_{\vk}^+ \Delta_{\vk}^-)^2 + M^2 ((\xi_{\vk}^+)^2 + (\Delta_{\vk}^+)^2)}
\end{align*}

\noindent where $\xi_\vk^\pm = (\xi_\vk \pm \xi_{\vk+\vQ})/2$ and $\Delta_\vk^\pm = (\Delta_\vk \pm \Delta_{\vk+\vQ})/2$. In the regions where the normal state FS is preserved perpendicular to $\vQ$, the eigenvalues reduce to the typical pure SC eigenvalues:

\begin{equation}
E_\vk^{(1)} = \sqrt{\xi_\vk^2 + \Delta_\vk^2}, \hspace{.25in} E_\vk^{(2)} = \sqrt{\xi_{\vk+\vQ}^2 + \Delta_{\vk+\vQ}^2}.
\end{equation}

\noindent When $T > T_c$ and $\Delta = 0$ the eigenvalues in the regions where the $\xi_\vk$ and $\xi_{\vk + \vQ}$ FSs overlap and the FS reconstructs reduce to the pure SDW eigenvalues: 

\begin{equation}
\begin{split}
   E_\vk^{(1)} = E_\vk^{(\alpha)} = \xi_\vk^+ + \sqrt{(\xi_\vk^-)^2 + M^2} \\ 
    E_\vk^{(2)} = E_\vk^{(\beta)} = \xi_\vk^+ - \sqrt{(\xi_\vk^-)^2 + M^2}.
\end{split}
\end{equation}

\noindent The eigenvalues in the region where the normal state FS is preserved reduce to $E^{(1)}_\vk = \xi_\vk$ and $E^{(2)}_\vk = \xi_{\vk + \vQ}$. Furthermore, the FS of this system is reconstructed from the black curves in FIG. ~\ref{fig:SDW_nesting} to the black curves in FIG. ~\ref{fig:SC_nodes} when $T < T_N$ where it can be seen that sections of the FS parallel to $\vQ$ become gapped by the SDW order.

\subsection{\label{sec:symmetry} Symmetry classes of the SC order parameters}

The coexistence of the SDW and SC order parameters $M$ and $\Delta$ depends on the symmetry of the SC order parameter translated by the SDW nesting vector. If translations of the SC order parameter by $\vQ$ are even ($\Delta_{\vk + \vQ} = \Delta_{\vk}$), denoted by $(E)$, then the order parameters are competitive with each other and the existence of SC order suppresses the SDW order and the SC transition temperature ($T_c$)\cite{kato}, which can be seen from the orange curves in FIG. ~\ref{fig:deltaM}. Whereas if the SC order parameter is odd under translations of the nesting vector ($\Delta_{\vk + \vQ} = - \Delta_{\vk}$), denoted by $(O)$, then the order parameters are cooperative with each other and the existence of SC order enhances the SDW on-site magnetization and $T_c$\cite{kato}, which can be seen from the cyan curves in FIG. ~\ref{fig:deltaM}.

\begin{figure}[t!]
\begin{centering}
\includegraphics[width=8.5cm]{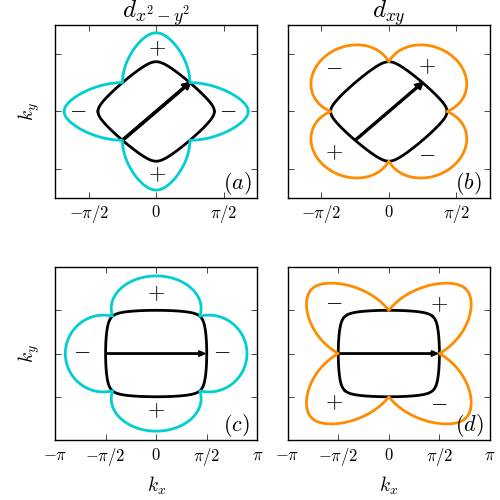}
\caption{Qualitative illustration of the amplitude and sign of the superconducting gap along the normal state tight-binding FSs (band parameters $t_1 = 100 T_N$ and $t_2 = 10 T_N$) to show the symmetry of the SC gap under translations of the $\vQ$-vector. (a) $d_{x^2-y^2}$ SC pairing on the $\xi_{\vk}^{(1)}$ FS (b) $d_{xy}$ pairing on the $\xi_{\vk}^{(1)}$ FS (c) $d_{x^2-y^2}$ SC pairing on the $\xi_{\vk}^{(2)}$ FS (d) $d_{xy}$ pairing on the $\xi_{\vk}^{(2)}$ FS.} 
\label{fig:SC_symm}
\end{centering}
\end{figure}

\begin{table}[b!]
\caption{\label{tab:table1} Symmetry class of the SC basis function,  $\mathcal{Y}_{\vk}$, under translations of the SDW nesting vector $\vQ$.}
\begin{ruledtabular}
\begin{tabular}{c c c}
$\vQ$ & $d^{(O)}$-wave & $d^{(E)}$-wave\\
\hline
$(\pi/2,\pi/2)$ & $d_{x^2-y^2}$ & $d_{xy}$\\
$(\pi,0)$ & $d_{xy}$ & $d_{x^2-y^2}$\\
\end{tabular}
\end{ruledtabular}
\end{table}

As can be seen from FIG. ~\ref{fig:SC_symm}, for the SDW state with nesting vector $\vQ = (\pi/2,\pi/2)$, the SC gap is even under translations of $\vQ$ for $d_{xy}$ and odd for $d_{x^2-y^2}$. However for the SDW state with $\vQ = (\pi,0)$ the $d_{x^2-y^2}$ SC pairing state is even under translations of $\vQ$ and $d_{xy}$ is odd under these same translations (these symmetry classifications are listed in Table ~\ref{tab:table1}). This switching between the $d$-wave symmetry classes under translations of $\vQ$ is a direct result of both the $\vQ$ vectors and the $d_{x^2-y^2}$ and $d_{xy}$ SC basis functions ($\mathcal{Y}_\vk$) being rotated by $\theta_\vk = \pi/4$ relative to each other. Therefore to maintain the same symmetry in $\Delta_\vk$ under translations of $\vQ$, the $d$-wave basis functions are switched when $\vQ$ is rotated.

\begin{figure*}[t!]
\begin{centering}
\includegraphics[width=7in]{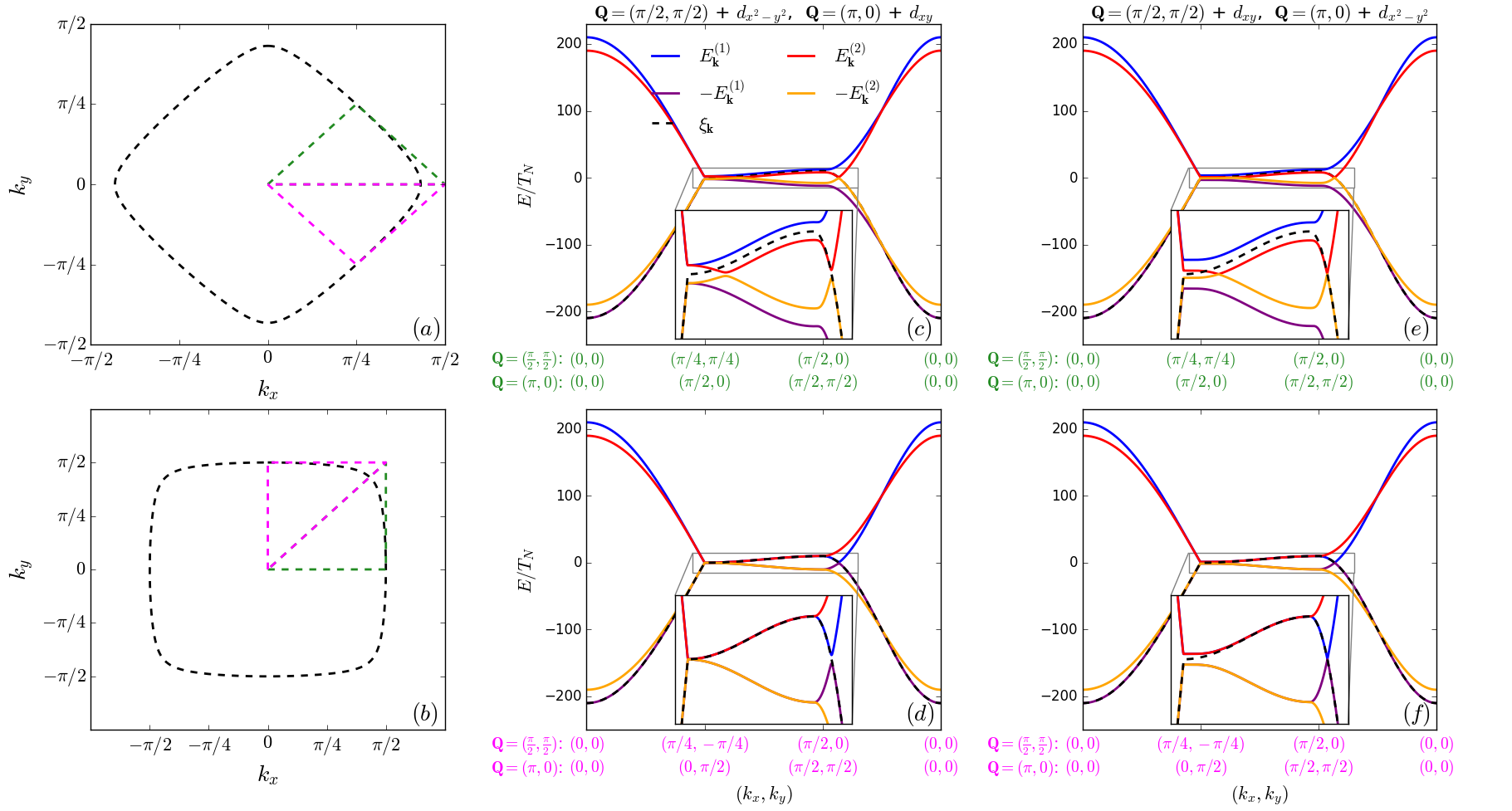}
\caption{(a) Band paths relative to the $\xi_\vk^{(1)}$ FS for the $\vQ = (\pi/2,\pi/2)$ SDW systems both parallel to $\vQ$ (green) and perpendicular to $\vQ$ (magenta). (b) Band paths relative to the $\xi_\vk^{(2)}$ FS for the $\vQ = (\pi,0)$ SDW systems both parallel to $\vQ$ (green) and perpendicular to $\vQ$ (magenta). (c) Band structure for the $d$-wave SC gap which is odd under translations of $\vQ$ plotted along the band paths parallel to $\vQ$ for both nesting vectors (green paths). (d) Band structure for the $d$-wave SC gap which is odd under translations of $\vQ$ plotted along the band paths perpendicular to $\vQ$ for both nesting vectors (magenta paths). (e) Band structure for the $d$-wave SC gap which is even under translations of $\vQ$ plotted along the band paths parallel to $\vQ$ for both nesting vectors (green paths). (f) Band structure for the $d$-wave SC gap which is even under translations of $\vQ$ plotted along the band paths perpendicular to $\vQ$ for both nesting vectors (magenta paths).} 
\label{fig:bandPlot}
\end{centering}
\end{figure*}

The nature of the zero-energy excitations critically depends on the symmetry of the SC gap under translations of the nesting vector\cite{senchoudhury}. When this symmetry is even, $\Delta_{\vk+\vQ} = \Delta_{\vk}$ so $\Delta_\vk^+ = \Delta_\vk$ and $\Delta_\vk^- = 0$. Similarly, when this symmetry is odd, $\Delta_{\vk+\vQ} = -\Delta_{\vk}$ so $\Delta_\vk^+ = 0$ and $\Delta_\vk^- = \Delta_\vk$. This simplifies the eigenvalues when both $\Delta$ and $M$ are nonzero in Equation (\ref{eq:coexistEigs}) to:

\begin{equation}
    E_\vk^{(1,2;E,O)} = \sqrt{\Gamma_\vk \pm 2\Lambda_\vk^{(E,O)}}
\end{equation}
\begin{align*}
    \Gamma_\vk &= (\xi_\vk^+)^2 + (\xi_\vk^-)^2 + \Delta_\vk^2 + M^2 \\ 
    \Lambda_\vk^{(E)} &= \sqrt{(\xi_\vk^+ \xi_\vk^-)^2 + M^2((\xi_\vk^+)^2+\Delta_\vk^2)} \\
    \Lambda_\vk^{(O)} &= \sqrt{(\xi_\vk^+ \xi_\vk^-)^2 + M^2(\xi_\vk^+)^2} \\
\end{align*}

\noindent For even symmetry, nodal points exist in addition to the SC symmetry nodes near the edge of the RBZ where $\xi_\vk^- = 0$ and when $E_\vk^{(2;E)} = M - \sqrt{(\xi_\vk^+)^2 + \Delta_\vk^2} = 0$. The locations of these nodes can be seen as the red crosses in FIG. ~\ref{fig:SC_nodes}. In the odd symmetry cases, these additional nodes are absent. Furthermore, two of the $d$-wave symmetry nodes with odd translational symmetry are gapped by the SDW order due to their nodal lines aligning with $\vQ$. This can be seen from the cyan $d_{x^2-y^2}$ and orange $d_{xy}$ nodal lines parallel to $\vQ$ in FIG. ~\ref{fig:SC_nodes} (a) and (b) respectively.

The resulting band structures are shown in FIG. ~\ref{fig:bandPlot} where the first two panels, FIG. ~\ref{fig:bandPlot} (a) and (b), illustrate the band paths relative to the normal state FSs. The middle two panels, FIG. ~\ref{fig:bandPlot} (c) and (d), display the calculated band structure for both SDW states coexisting with the $d$-wave SC gaps which are odd under translations of $\vQ$ along the indicated paths. The final two panels, FIG. ~\ref{fig:bandPlot} (e) and (f), display the calculated band structure for both SDW states coexisting with the $d$-wave SC gaps which are even under translations of $\vQ$ again along the indicated paths. For $\vQ = (\pi/2,\pi/2)$ the odd $d$-wave state is the $d_{x^2-y^2}$ pairing and for $\vQ = (\pi,0)$ the odd state is the $d_{xy}$ pairing state. In FIG. ~\ref{fig:bandPlot}(c), the bands shown are those in the region where SDW ordering reconstructs the FS parallel to $\vQ$ along the green paths. The SC node typically present in the vicinity of $\vk = (\pi/4,\pi/4)$ [for $\vQ = (\pi/2,\pi/2)$] or near $\vk = (\pi/2,0)$ [for $\vQ = (\pi,0)$] becomes gapped by the SDW. This again corresponds to the cyan $d_{x^2-y^2}$ and orange $d_{xy}$ nodal lines in FIG. ~\ref{fig:SC_nodes} (a) and (b), respectively. In FIG. ~\ref{fig:bandPlot}(d) the bands determined along the magenta paths in FIG. ~\ref{fig:bandPlot} (a) and (b) are shown where the normal state FS is preserved perpendicular to $\vQ$, resulting in the pure SC bands and the $d$-wave symmetry nodes in the vicinity of $\vk = (\pi/4,-\pi/4)$ [for $\vQ = (\pi/2,\pi/2)$] or near $\vk = (0,\pi/2)$ [for $\vQ = (\pi,0)$] are preserved.

In FIG. ~\ref{fig:bandPlot} (e) and (f) the band structure is shown for the SDW states coexisting with the $d$-wave SC gaps with even symmetry when translated by $\vQ$. For the $\vQ = (\pi/2,\pi/2)$ SDW state this is the $d_{xy}$ pairing state and for the $\vQ = (\pi,0)$ SDW state this is the $d_{x^2-y^2}$ pairing state. The quasiparticle bands parallel to $\vQ$ along the green paths in FIG. ~\ref{fig:bandPlot} (a) and (b) are shown in FIG. ~\ref{fig:bandPlot}(e). The typical $d$-wave nodes occurring in the vicinity of $\vk = (\pi/2,0)$ [for $\vQ = (\pi/2,\pi/2)$] or $\vk = (\pi/2,\pi/2)$ [for $\vQ = (\pi,0)$] remain intact when both SC and SDW coexist. In addition to the $d$-wave symmetry nodes, additional nodes appear parallel to $\vQ$ near $\vk = (\pi/4,\pi/4)$ [when $\vQ = (\pi/2,\pi/2)$] or near $\vk = (\pi/2,0)$ [when $\vQ = (\pi,0)$]. These additional nodes correspond to the red crosses in FIG. ~\ref{fig:SC_nodes}. FIG. ~\ref{fig:bandPlot}(f) displays the band structure along the magenta band paths in FIG. ~\ref{fig:bandPlot} (a) and (b) where the normal state FS is preserved, again resulting in the pure SC bands. 

\subsection{\label{sec:conductivity} Kinetic method for heat conductivity}

From these band structures, the electronic thermal conductivity was calculated using the Boltzmann kinetic equation similar to calculations for the thermal conductivities of both $s$-wave and unconventional superconductors. Within the Boltzmann kinetic approach the thermal conductivity tensor is given by the equation\cite{mineev}

\begin{equation}
    \kappa_{ij} = -\frac{2}{T} \sum_{n = 1}^2 \int \frac{d^2k}{(2\pi)^2} (E^{(n)}_\vk)^2 (\mathbf{v}_\vk^{(n)})_i (\mathbf{v}_\vk^{(n)})_j \frac{\partial f^0_\vk}{\partial E} (\tau_{n1}^{-1} + \tau_{n2}^{-1})^{-1}
    \label{eq:conductivity}
\end{equation}

\noindent where $f^0_\vk = \frac{1}{e^{E_\vk/T} + 1}$ is the equilibrium Fermi-Dirac distribution function, $\mathbf{v}_\vk^{(n)} = \mathbf{\nabla}_\vk E^{(n)}_\vk$ is the quasiparticle velocity, and $\tau_{nm}$ is the quasiparticle relaxation time defined as

\begin{equation}
    \tau_{nm}^{-1}(\vk) = N_{\text{imp}} V^2 \frac{2\pi}{\hbar} \int \frac{d^2k'}{(2\pi)^2} |C_{nm}(\vk,\vk')|^2 \delta(E_\vk^{(n)} - E_{\vk'}^{(m)})
    \label{eq:scattering}
\end{equation}

\noindent where $C_{nm}(\vk,\vk')$ is known as the coherence factor and is the amplitude for a single impurity to scatter a quasiparticle from the state with momentum $\vk$ and energy $E^{(n)}_\vk$ to a state with momentum $\vk'$ and energy $E^{(m)}_{\vk'}$ within the Born limit\cite{senchoudhury}. $N_{\text{imp}}$ is the density of impurities, and $V$ is the isotropic scattering amplitude, where $N_{\text{imp}}V \ll 1$ in the limit of weak impurity scattering. The quasiparticle relaxation time integral was calculated numerically with the unknown $N_{\text{imp}} V^2$ eliminated in favor of the normal state quasiparticle relaxation times; $\tau_{\text{n}}^{-1} = N_{\text{imp}} V^2 \frac{2\pi}{\hbar}N_F$, where $N_F$ is the density of states at the Fermi level in the normal state. Furthermore, $\tau_{\text{n}}^{-1}$ cancels out for the choice of normalization used in this work, $\kappa(T)/\kappa(T_N)$.

The coherence factors can be calculated from the impurity scattering Hamiltonian:

\begin{equation}
\begin{split}
    H_{\text{imp}} &= V \sum_{\vk,\vk',\sigma} \hat{a}_{\vk',\sigma}^\dagger \hat{a}_{\vk,\sigma} \\
    &= \sum_{\vk,\vk',\sigma} \big(\hat{\mathbf{\Psi}}_{\vk'}^{\text{n}}\big)^\dagger \hat{\mathcal{H}}^{\text{imp}}_\vk \hat{\mathbf{\Psi}}_{\vk}^{\text{n}}
\end{split}
\end{equation}

\noindent where $\hat{\mathcal{H}}^{\text{imp}}_\vk$ is the impurity scattering Hamiltonian in the Nambu basis, and can be written as:

\begin{equation}
    \hat{\mathcal{H}}^{\text{imp}}_\vk = \frac{V}{4} \begin{pmatrix}
    1 & 0 & 0 & 0 \\
    0 & -1 & 0 & 0 \\
    0 & 0 & 1 & 0 \\
    0 & 0 & 0 & -1 \\
    \end{pmatrix}
\end{equation}

\noindent which can be rewritten in the basis of the coexistence state Nambu vector to reveal the matrix of coherence factors:

\begin{equation}
    H_{\text{imp}} = \sum_{\vk,\vk',\sigma} \hat{\mathbf{\Psi}}_{\vk'}^\dagger \hat{D}(\vk,\vk') \hat{\mathbf{\Psi}}_{\vk}
\end{equation}

\noindent where $\hat{\mathbf{\Psi}}_\vk^\dagger = (\hat{c}_{1,\vk,\sigma}^\dagger,\hat{c}_{1,-\vk,-\sigma},\hat{c}_{2,\vk,\sigma}^\dagger,\hat{c}_{2,-\vk,-\sigma})$ is the Nambu state vector for the coexistence quasiparticle bands (this can be generalized to accommodate Nambu vectors for the SDW and SC quasiparticles since $M$ and $\Delta$ aren't always nonzero depending on $T$ and $\vk$). Quasiparticles occupying states in the $E^{(1)}_\vk$ band with momentum $\vk$ are defined by $\hat{c}_{1,\vk,\sigma}^\dagger \ket{0}$ and quasiparticles occupying states in the $E^{(2)}_\vk$ band with momentum $\vk$ are defined by $\hat{c}_{2,\vk,\sigma}^\dagger \ket{0}$, where $\ket{0}$ is the vacuum state with no quasiparticles. Performing the Bogoliubov transformation on the impurity scattering Hamiltonian yields the matrix of coherence factors for the quasiparticle and quasihole bands, $\hat{D}(\vk,\vk')$:

\begin{equation}
    \hat{D}(\vk,\vk') = \hat{B}_{\vk'} \hat{\mathcal{H}}^{\text{imp}}_\vk \hat{B}_{\vk}^\dagger
    \label{eq:coherence}
\end{equation}

\noindent where $\hat{B}_\vk$ is the Bogoliubov transformation matrix, the structure of which depends on whether or not $\Delta$ and/or $M$ is nonzero and the symmetry class of the SC gap function. The details of the Bogoliubov transformation matrices have been worked out in Appendix ~\ref{sec:twoStepDiag}. The intraband quasiparticle band coherence factors are:
\begin{equation}
    C_{11}(\vk,\vk') = D_{11}(\vk,\vk'), \hspace{.1in} C_{22}(\vk,\vk') = D_{33}(\vk,\vk')
\end{equation}
\noindent and the interband quasiparticle band  coherence factors are: 
\begin{equation}
    C_{12}(\vk,\vk') = D_{13}(\vk,\vk'), \hspace{.1in} C_{21}(\vk,\vk') = D_{31}(\vk,\vk')
\end{equation}

\noindent where the $-E^{(1)}_\vk$ and $-E^{(2)}_\vk$ bands have been neglected due to quasiparticle-quasihole symmetry in the model. The calculation of the coherence factors from $\hat{\mathcal{H}}^{\text{imp}}_\vk$ and the Bogoliubov transformation matrices was performed numerically.

A more simple case to consider analytically is that of superconductivity in the absence of a coexistence state, such as spin density waves. In the Born limit, the coherence factor is known\cite{mineev} to be:

\begin{equation}
    \abs{C^{\text{SC}}(\vk,\vk')}^2 = \frac{1}{2} \bigg(1+\frac{\xi_\vk \xi_{\vk'} - \Delta_\vk \Delta_{\vk'}}{E_\vk E_{\vk'}} \bigg)
\end{equation}

\noindent where the quasiparticle energy for a superconductor is defined as $E_\vk = \sqrt{\xi_\vk^2 + \Delta_\vk^2}$. The $\xi_\vk \xi_{\vk'}$ term in this coherence factor integrates to $0$ by symmetry in Equation (\ref{eq:scattering}). For a $d$-wave SC gap, the $\Delta_\vk \Delta_{\vk'}$ term also integrates to $0$ due to having symmetric positive and negative $\Delta_\vk$ values on the bare tight-binding FS. Thus, the quasiparticle lifetimes of the $d$-wave state in the Born limit is inversely proportional to the DOS of the superconducting quasiparticle states\cite{mineev}$^,$\cite{lee}, $N(E_\vk)$; $\tau^{d}_\vk = \tau_n N_F / N(E_\vk)$. However, on symmetry broken tight-binding FSs this term doesn't necessarily integrate to $0$, as was the case when integrated on a distorted FS due to nematicity\cite{senchoudhury2}. Another case which can be discussed analytically is that of the SDW state in the absence of superconductivity. The intraband coherence factors of this state can be written\cite{senchoudhury} as:

\begin{equation}
    \abs{C_{\text{11}}^{\text{SDW}}(\vk,\vk')}^2 = \abs{C_{\text{22}}^{\text{SDW}}(\vk,\vk')}^2 = \frac{1}{2}\bigg(1 + \frac{\xi_\vk^- \xi_{\vk'}^- + M^2}{\zeta_\vk \zeta_{\vk'}} \bigg)
\end{equation}

\noindent and the interband coherence factors can be written\cite{senchoudhury} as:

\begin{equation}
    \abs{C_{\text{12}}^{\text{SDW}}(\vk,\vk')}^2 = \abs{C_{\text{21}}^{\text{SDW}}(\vk,\vk')}^2 = \frac{1}{2}\bigg(1 - \frac{\xi_\vk^- \xi_{\vk'}^- + M^2}{\zeta_\vk \zeta_{\vk'}} \bigg)
\end{equation}

\noindent where $\zeta_\vk = \sqrt{(\xi_\vk^-)^2 + M^2}$. While in the limit of a perfectly nested SDW state ($t_2 = 0$), the $\xi_\vk^- \xi_{\vk'}^-$ can be shown to integrate to $0$ and the quasiparticle lifetimes become $\tau^{\text{SDW}}_{11}= \tau_n N_F / N(E^{(\alpha)}_\vk) (1+M^2/(E^{(\alpha)}_\vk)^2)^{-1}$, $\tau^{\text{SDW}}_{22} = \tau_n N_F / N(E^{(\beta)}_\vk) (1+M^2/(E^{(\beta)}_\vk)^2)^{-1}$, and $\tau^{\text{SDW}}_{12} = \tau^{\text{SDW}}_{21} = 0$\cite{senchoudhury}, however such symmetry arguments cannot be made away from perfect nesting and the lifetimes need to be calculated numerically.

\section{Numerical Results and Discussion}

The Cartesian components of the thermal conductivity tensor $\kappa_{xx}(T)$, $\kappa_{yy}(T)$, and $\kappa_{xy}(T)$ ($\kappa_{yx}(T) = \kappa_{xy}(T)$) were numerically calculated in the RBZ. However, the frame-of-reference of the nesting vector $\vQ$ diagonalizes the thermal conductivity tensor, and is therefore the more natural frame to study thermal transport. This is straight-forwardly accomplished by rotating the coordinate system by $\theta = \pi/4$ for the case when $\vQ = (\pi/2,\pi/2)$. The rotated conductivity tensor $\hat{\kappa}' = \hat{R}(\theta = \pi/4) \hat{\kappa} \hat{R}^T(\theta = \pi/4)$ is:

\begin{equation}
    \begin{pmatrix}
    \kappa_\perp & 0 \\
    0 & \kappa_\parallel \\
    \end{pmatrix} = \frac{1}{2}\begin{pmatrix}
    1 & -1 \\
    1 & 1 \\
    \end{pmatrix}
    \begin{pmatrix}
    \kappa_{xx} & \kappa_{xy} \\
    \kappa_{xy} & \kappa_{xx} \\
    \end{pmatrix}
    \begin{pmatrix}
    1 & 1 \\
    -1 & 1 \\
    \end{pmatrix}
\end{equation}

\noindent which was simplified by using the inherent symmetries ($\kappa_{xx} = \kappa_{yy}$, $\kappa_{yx} = \kappa_{xy}$) in the case when $\vQ = (\pi/2,\pi/2)$, and leads to the diagonalized thermal conductivity components $\kappa_\perp = \kappa_{xx} - \kappa_{xy}$ and $\kappa_\parallel = \kappa_{xx} + \kappa_{xy}$. In the SDW system with $\vQ = (\pi,0)$, no rotation is needed and $\kappa_{xy}$ integrates to 0, resulting in a diagonal thermal conductivity tensor. 

\begin{figure}[t!]
\begin{centering}
\includegraphics[width=8.5cm]{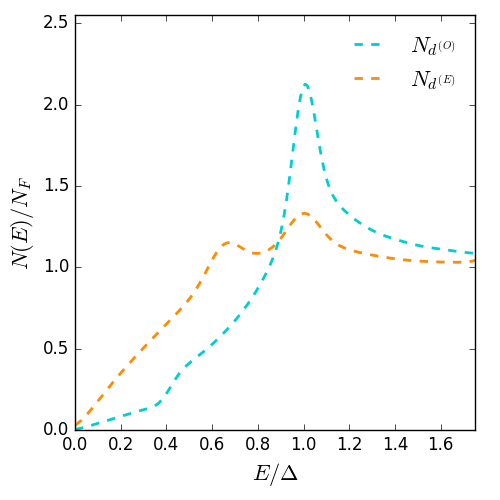}
\caption{Density of states normalized by the normal state DOS at the Fermi level ($N_F$) for SDW + $d$-wave SC order with odd symmetry under translations of $\vQ$ (cyan) and SDW + $d$-wave SC order with even symmetry under translations of $\vQ$ (orange) plotted vs. energy normalized by the SC gap maxima on the FS ($\Delta$).} 
\label{fig:DOS}
\end{centering}
\end{figure}

Appropriately integrating the band structure generates the electronic density of states (DOS) for both SC + SDW coexistence states. One important aspect of the nodal structures is the variation in the DOS just above the Fermi level (which occurs when $E = 0$) for the two SC + SDW coexistence states (shown in FIG. ~\ref{fig:DOS}). The enhancement occurring just above the Fermi level has important consequences for the low-$T$ behavior of the thermal conductivity elements.

The electronic thermal conductivity calculated on the normal state tight-binding FS was found to have a linear dependence on temperature. To accentuate the deviation from the normal state thermal conductivity, the conductivity elements in FIG. ~\ref{fig:kappa_pureSC} and FIG. ~\ref{fig:thermal_conductivity}(a) had their linear $T$-dependence removed by plotting $\kappa(T)/T$ and were normalized by $\kappa(T_N)/T_N$. In FIG. ~\ref{fig:thermal_conductivity}(b) the conductivity elements were normalized by $\kappa(T_N)$ on a log-log scale to emphasize which conductivity elements preserve this linear $T$-dependence and which ones deviate from it as $T \rightarrow 0$. 

\subsection{\label{sec_pureSC_kappa} Pure $d$-wave SC Thermal Conductivity}

\begin{figure}[h!]
\begin{centering}
\includegraphics[width=8.5cm]{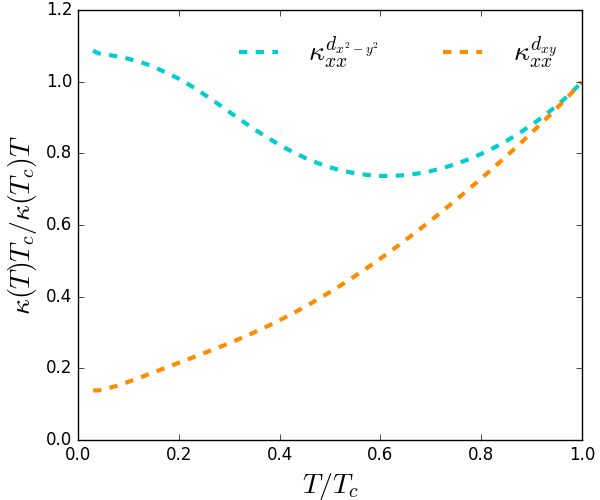}
\caption{Thermal conductivity $\kappa_{xx}(T)/T$ calculated for a tight-binding dispersion $\xi_\vk = \mu - t_1\cos{k_x} - t_1\cos{k_y} - t_2\cos{k_x}\cos{k_y}$ (where $\mu = 0$, $t_1 = 100 T_c$, and $t_2 = 10 T_c$) with $d_{x^2-y^2}$ (cyan) and $d_{xy}$ (orange) SC gaps normalized by $\kappa_{xx}(T_c)/T_c$ to remove the $T$-linearity.} 
\label{fig:kappa_pureSC}
\end{centering}
\end{figure}

The thermal conductivity of a $d$-wave superconductor in the absence of a coexistence state on a tight-binding electronic dispersion ($\xi_\vk = \mu - t_1\cos{k_x} - t_1\cos{k_y} - t_2\cos{k_x}\cos{k_y}$) has been previously calculated in literature\cite{senchoudhury2}$^,$\cite{senchoudhury}. This has also been calculated in this work (see FIG. ~\ref{fig:kappa_pureSC}) to compare the thermal conductivity of the SC + SDW coexistence states to. Since the $C(4)$ rotational symmetry of the tight-binding FS is preserved for the pure superconducting state, $\kappa_{xx} = \kappa_{yy}$ and $\kappa_{xy} = \kappa_{yx} = 0$ by symmetry.

For $d$-wave superconductors the quasiparticle lifetimes can be shown to only depend inversely on the quasiparticle DOS\cite{mineev}$^,$\cite{lee}
($\tau^d_\vk = \tau_n N_F / N(E_\vk)$). The quasiparticle lifetimes for the $d_{x^2-y^2}$ and $d_{xy}$ states are very different due to the difference in the local DOS of their respective nodal quasiparticle states. The difference in the local DOS at the nodes of these $d$-wave SC states is due to the curvature of the FS where these nodes occur; flat regions of the FS result in a low local DOS and curved regions of the FS result in a high local DOS. The $d_{x^2-y^2}$ SC gap has nodal quasiparticle states with a low local DOS, and therefore longer lifetimes, and high Fermi velocities. Whereas, the $d_{xy}$ SC gap leads to nodal quasiparticle states with a high local DOS, and therefore short lifetimes, and low Fermi velocities. Therefore, the low-$T$ thermal transport of these $d$-wave states is profoundly different, as the $d_{x^2-y^2}$ SC gap has nodal quasiparticles which are both long-lived and high-velocity, while the $d_{xy}$ SC gap has nodal quasiparticles which are both short-lived and low-velocity. This leads to the low-$T$ thermal conductivity for the $d_{x^2-y^2}$ SC gap being slightly enhanced relative to the normal state, and the low-$T$ thermal conductivity for a $d_{xy}$ SC gap being greatly diminished relative to the normal state (see FIG. ~\ref{fig:kappa_pureSC}).

\subsection{\label{sec:SDW_kappa} Pure SDW Thermal Conductivity}

\begin{figure}[t!]
\begin{centering}
\includegraphics[width=8.5cm]{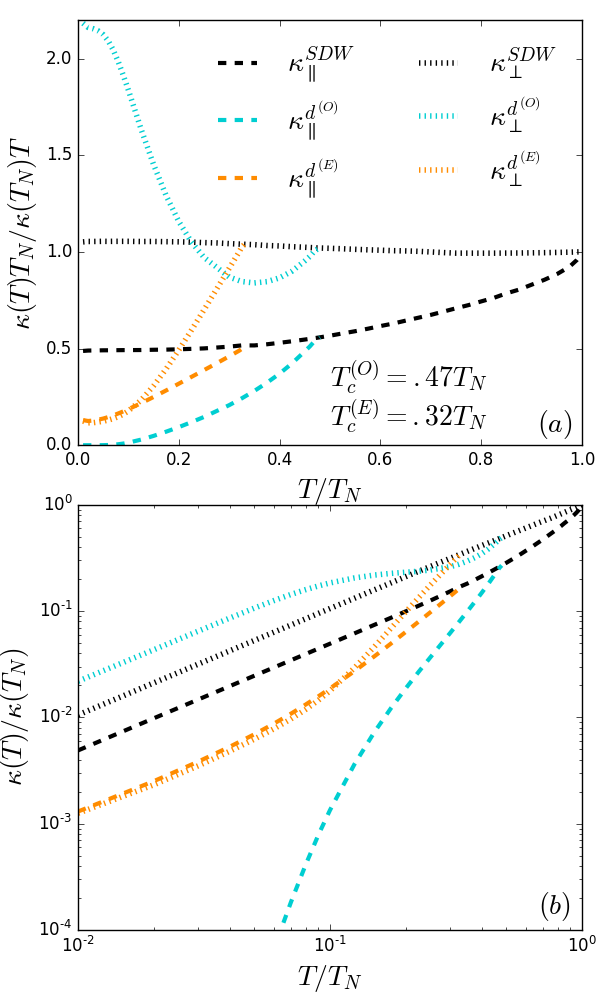}
\caption{Diagonalized thermal conductivity tensor elements parallel (dashed curves) and perpendicular (dotted curves) to the SDW nesting vector $\vQ$ in the pure SDW state (black), SDW + $d$-wave SC state with odd translational symmetry (cyan), and SDW + $d$-wave SC state with even translational symmetry (orange). (a) $\kappa(T) T_N/\kappa(T_N) T$ plotted to remove linear $T$- dependence and emphasize deviations from normal state conductivity. (b) $\kappa(T)/\kappa(T_N)$ plotted on a log-log scale to emphasize low-$T$ linearity.} 
\label{fig:thermal_conductivity}
\end{centering}
\end{figure}

In the pure SDW case the electronic thermal conductivities were calculated numerically in the directions parallel and perpendicular to the nesting vector $\vQ$ (see the black curves of FIG. ~\ref{fig:thermal_conductivity}). Parallel to the nesting vector, the thermal conductivity $\kappa_{\parallel}^{\text{SDW}}(T)$ falls sharply when compared to that of the normal state, which is often seen in thermal conductivity measurements of spin density wave antiferromagnets\cite{kim}$^,$\cite{steckel}$^,$\cite{sayles}. This fall in the conductivity can be attributed to a growing gap in the reconstructed FS as $T$ decreases from $T_N$, which can be seen in FIG. ~\ref{fig:SC_nodes}. As the on-site magnetization, $M(T)$, reaches its maximum value, $\kappa_{\parallel}^{SDW}(T)/T$ becomes a constant, diminished from the normal state conductivity. This result is comparable to the pure SDW thermal conductivity for the case when $\vQ = (\pi,\pi)$ with a nearly half electron filling for similar band parameters\cite{senchoudhury}. Perpendicular to $\vQ$ in these SDW systems the normal state tight-binding FS is preserved and $\kappa_{\perp}^{SDW}(T)$ is essentially that of the normal state thermal conductivity. However, as $T \rightarrow 0$ this thermal conductivity becomes slightly enhanced due to the gap in the FS generated by the SDW order which reduces the available states at the Fermi level, increasing the quasiparticle lifetimes resulting in a slightly enhanced thermal conductivity.

\subsection{\label{sec:d_kappa} SDW + $d$-wave SC Thermal Conductivity}

When the propagation direction of the SDW is aligned (parallel case) with a $d$-wave nodal line, the FS reconstruction destroys two of the $d$-wave symmetry nodes (e.g. this occurs for the $\vQ = (\pi/2,\pi/2)$ SDW and the $d_{x^2-y^2}$ SC gap or $\vQ = (\pi,0)$ SDW and the $d_{xy}$ SC gap), while the symmetry nodes perpendicular to the SDW propagation direction are unaffected. This can be seen by the cyan lines in FIG. ~\ref{fig:SC_nodes}(a) and the orange lines in FIG. ~\ref{fig:SC_nodes}(b), where the remaining nodes occur on the low local DOS regions (flat regions) of the FS. These two coexistence phases have SC gaps which are odd under translations of their respective $\vQ$ vectors (i.e. $\Delta_{\vk + \vQ} = -\Delta_\vk$) (see FIG. ~\ref{fig:SC_symm} (a) and (d)). More so, these systems have equivalent band structures along their respective band paths (see FIG. ~\ref{fig:bandPlot} (c) and (d)). Due to these similarities these coexistence states have equivalent transport properties relative to their $\vQ$ vectors, which will be referred to as the $d^{(O)}$-wave state (odd symmetry state).

The $\vQ = (\pi/2,\pi/2)$ SDW state coexisting with the $d_{xy}$ SC gap and the $\vQ = (\pi,0)$ SDW state coexisting with the $d_{x^2-y^2}$ SC gap have symmetry nodes on the high local DOS regions (curved regions) of their respective normal state FSs, all of which remain unchanged by the FS reconstruction (note the orange points in FIG. ~\ref{fig:SC_nodes}(a) and the cyan points in FIG. ~\ref{fig:SC_nodes}(b)). These coexistence states both have SC gaps which are even under translations of $\vQ$ (see FIG. ~\ref{fig:SC_symm} (b) and (c)). Due to this translational symmetry, additional mixing nodes appear near the FS reconstruction which can be seen as the red crosses in FIG. ~\ref{fig:SC_nodes}. Furthermore, these coexistence states have equivalent band structures along their respective band paths (see FIG. ~\ref{fig:bandPlot} (e) and (f)) which lead to them having equivalent transport properties relative to $\vQ$. These states will be referred to as the $d^{(E)}$-wave state (even symmetry state).

\subsubsection{\label{sec:dx2y2_kappa} SDW + $d^{(O)}$-wave SC Thermal Conductivity}

Perpendicular to $\vQ$, this system behaves like a pure $d$-wave SC system, similar to the previously discussed $d_{x^2-y^2}$ SC gap on a tight-binding FS. However, only half the symmetry nodes typically present in similar $d$-wave SC systems survive the FS reconstruction, so the DOS just above the Fermi level ($E = 0$) is approximately half that of the pure $d$-wave system with the same band parameters (note the reduction in the DOS just above the Fermi level for $d^{(O)}$ in FIG. ~\ref{fig:DOS}). With half the available states to scatter to just above the Fermi level, there is a doubling in the quasiparticle lifetimes at the remaining nodes relative to the lifetimes of quasiparticles occupying nodal states in similar $d$-wave superconductors without coexisting spin density waves. Since the remaining symmetry nodes contain quasiparticles with Fermi velocities purely perpendicular to $\vQ$ with lifetimes approximately twice that of their pure SC counterparts, the residual $\kappa_{\perp}(T \rightarrow 0)/T$ is therefore roughly twice the residual $\kappa(T \rightarrow 0)/T$ for a pure $d$-wave SC with nodes located on the flat parts of a tight-binding FS\cite{senchoudhury} (compare cyan curves in FIG. ~\ref{fig:kappa_pureSC} to FIG. ~\ref{fig:thermal_conductivity}(a)). The thermal conductivity perpendicular to $\vQ$ is linear in $T$, (see FIG. ~\ref{fig:thermal_conductivity}(b)), and therefore behaves like a SC with line nodes in this direction\cite{graf}. The thermal conductivity parallel to $\vQ$ decreases exponentially as $T \rightarrow 0$ since the nodal quasiparticle states which would have Fermi velocities parallel to $\vQ$ for SCs with similar nodal lines have been gapped by the FS reconstruction, and therefore the system behaves like a fully-gapped SC in this direction\cite{bardeen}.

\subsubsection{\label{sec:dxy_kappa} SDW + $d^{(E)}$-wave SC Thermal Conductivity}

Since none of these $d$-wave symmetry nodes become gapped by the FS reconstruction and additional mixing nodes appear due to the SC gap being even under translations of $\vQ$, the DOS just above the Fermi level for $d^{(E)}$ is relatively large (see FIG. ~\ref{fig:DOS}) and results in short-lived quasiparticles. The majority of the states just above the Fermi level are located in the vicinity of the $d$-wave symmetry nodes rather than the mixing nodes, which results in the symmetry nodes dominating low-$T$ thermal transport. Quasiparticles occupying states at the symmetry nodes have Fermi velocities with equal magnitude components perpendicular and parallel to $\vQ$ and contribute equally to the low-$T$ transport in both directions. However, quasiparticles occupying states at the mixing nodes have Fermi velocities parallel to $\vQ$ and only contribute to $\kappa_{\parallel}$, thus leading to a system with weakly anisotropic thermal transport where $\kappa_{\parallel}(T \rightarrow 0) > \kappa_{\perp}(T \rightarrow 0)$ (see the orange curves in Fig 6 (a)). 

The residual $\kappa(T \rightarrow 0)/T$ thermal conductivity elements in these cases are much smaller than $\kappa_{\perp}(T \rightarrow 0)/T$ in the previously discussed $d$-wave system due to the quasiparticles at these $d$-wave symmetry nodes having significantly lower Fermi velocities and lifetimes. Furthermore, these residual $\kappa(T \rightarrow 0)/T$ values both parallel and perpendicular to $\vQ$ are nearly identical to those for the similar $d_{xy}$ SC gap on a tight-binding FS (compare orange curves in FIG. ~\ref{fig:kappa_pureSC} to FIG. ~\ref{fig:thermal_conductivity}(a)). This is due to the fact that the $d$-wave symmetry nodes for the $d^{(E)}$-wave case are largely unaffected by their coexistence with the SDW state, but this does introduce additional mixing nodes which slightly enhance $\kappa_\parallel(T)/T$ above the pure SC value. While the quasiparticles occupying the states at the mixing nodes have high Fermi velocities parallel to $\vQ$, the relative dearth of available states means they don't play a significant role in thermal transport. This $d$-wave SC state decreases linearly with $T$ at low $T$ as can be seen in FIG. ~\ref{fig:thermal_conductivity}(b) both parallel and perpendicular to $\vQ$, and therefore behaves like a SC with line nodes in both directions\cite{graf}.

%~~~~~~~~~~~~~~~~~~~~~~~~~~~~~~~~~~~~~~~~~~~~~~~~~~~~~~~~~~~~~~~~~~~~~~~~

\section{Conclusion} 

While this work ignores the effects of band multiplicity, these results are still useful in determining the nodal structures of commensurate SDW systems with nesting vectors $\vQ = (\pi/2,\pi/2)$ and $\vQ = (\pi,0)$ coexisting with singlet $d$-wave SC pairings. Commensurate SDW systems of type AF3 and AF2 were considered on two-dimensional tight-binding square-lattices and found to have equivalent transport properties within Boltzmann kinetic theory in the weak impurity scattering (Born) limit relative to their nesting vectors. Parallel to their nesting vectors these systems behave similar to a suppressed metal, where the electronic thermal conductivity is linear in $T$ but diminished from the normal state thermal conductivity. However, perpendicular to their nesting vectors, the transport properties of these systems are almost identical to that of the normal metallic state, except they are slightly enhanced as $T \rightarrow 0$ due to the FS reconstruction parallel to $\vQ$ creating a gap in the FS and reducing $N_F$, thus enhancing the quasiparticle lifetimes. 

The $d$-wave SC states coexisting with the $\vQ = (\pi/2,\pi/2)$ and $\vQ = (\pi,0)$ nesting vectors have equivalent transport properties, with the $d_{x^2-y^2}$ and $d_{xy}$ states swapped between the nesting vectors. The $d$-wave symmetry nodes are located on regions of the tight-binding FS with the same relatively small local DOS and have equivalent band structures when the $\vQ = (\pi/2,\pi/2)$ SDW state coexists with the $d_{x^2-y^2}$ SC gap and the $\vQ = (\pi,0)$ SDW state coexists with the $d_{xy}$ SC gap (these are the odd symmetry, $d^{(O)}$-wave state in this work). Similarly, the $d$-wave nodes occur on regions of the tight-binding FS with the same relatively large local DOS and have equivalent band structures when the $\vQ = (\pi/2,\pi/2)$ SDW state coexists with the $d_{xy}$ SC gap and the $\vQ = (\pi,0)$ SDW state coexists with the $d_{x^2-y^2}$ SC gap (these are the even symmetry, $d^{(E)}$-wave state in this work).

The electron transport properties of these commensurate SDW systems were also studied when SC singlet pairing arises ($T_N > T_c$) out of it. The electronic thermal conductivity for a $d^{(O)}$-wave SC gap measured parallel to $\vQ$ was found to decrease exponentially with $T$, consistent with results for fully-gapped SCs\cite{bardeen}. Perpendicular to $\vQ$, the conductivity was found to decrease linearly with $T$, consistent with SCs with line nodes\cite{graf}. Furthermore, the residual $\kappa_{\perp}(T \rightarrow 0)/T$ value was found to be roughly twice that of pure $d$-wave SCs containing nodal quasiparticle states situated on the flat regions of a tight-binding FS. Therefore, the effect SDW states with $C(2)$ rotational symmetry on $d$-wave SC states such as these is that it gaps the nodal quasiparticle states in the direction of SDW propagation, greatly reducing thermal transport in that direction, and doubles the lifetimes of quasiparticles traveling perpendicular to the SDW propagation direction, thus greatly enhancing thermal transport in that direction. This results in a system with highly anisotropic electron transport where fast long-lived quasiparticles tend to travel perpendicular to $\vQ$.

When the $d$-wave SC gap is even under translations of $\vQ$ ($d^{(E)}$-wave), the SC symmetry nodes are preserved as none of them appear in the region of the FS which is reconstructed by the SDW order. Due to the translational symmetry of the SC gap, additional mixing nodes appear in the vicinity of the FS reconstruction parallel to $\vQ$. The electronic thermal conductivity both parallel and perpendicular to $\vQ$ were found to decrease linearly with $T$. In fact, $\kappa_{\parallel}(T \rightarrow 0)/T$ and $\kappa_{\perp}(T \rightarrow 0)/T$ were nearly identical due to quasiparticles occupying states at the $d$-wave symmetry nodes contributing equally to thermal transport in both directions. However, thermal transport parallel to $\vQ$ was slightly enahnced since quasiparticles occupying states at the mixing nodes enhanced transport in that direction. The coexistence of SDW states with $C(2)$ $d$-wave SC gaps such as these leaves thermal transport of such systems largely unaffected due to the nodal quasiparticle states remaining mostly unchanged by the FS reconstruction, however it does introduce additional mixing nodes which slightly enhance thermal transport in the direction of SDW propagation. Therefore, this results in a system with weakly anisotropic thermal transport, where slow short-lived quasiparticles travel both parallel and perpendicular to $\vQ$, but slightly prefer to travel parallel to $\vQ$.

These results could be relevant in determining the nature of the $d$-wave gap in cuprates with commensurate SDW orders of nesting vectors: $\vQ = (\pi/2,\pi/2)$ or $\vQ = (\pi,0)$. If the thermal conductivity is measured both parallel and perpendicular to $\vQ$ and the nesting vector is known, weakly or strongly anisotropic thermal transport at low-$T$ could be used to determine whether the SC gap is $d_{x^2-y^2}$ or $d_{xy}$ in nature. Additionally, these results could be relevant to understanding the nature of anisotropic in-plane electronic thermal transport measured in some cuprate samples\cite{cohn}$^,$\cite{sun}$^,$\cite{sologubenko}. One study found quasi-one-dimensional electronic thermal transport at low-$T$ mediated by spin excitations\cite{sologubenko}, similar to the $d^{(O)}$-wave result in this work which had a residual thermal conductivity perpendicular to $\vQ$ as $T \rightarrow 0$, but not parallel to $\vQ$. However, the other studies\cite{cohn}$^,$\cite{sun} found that electronic thermal transport was supported in both directions, but still favored a particular direction due to electronic inhomogeneities. This effect could either be due to the fact that these samples weren't monolayers and the SDW nesting vectors for each layer weren't all parallel, or the anisotropic transport in these materials is due to a nematic phase which was discussed in a previous work\cite{senchoudhury2}.

\section*{Acknowledgements}

This work supported by NSF Award No. 1809846 and through the NSF MonArk Quantum Foundry supported under Award No. 1906383. The authors would like to thank Anton Vorontsov for the initial suggestion of the problem.

\appendix

\section{Order Parameter Self-Consistency}\label{sec:OPselfConsistency}

\begin{figure}[b!]
\begin{centering}
\includegraphics[width=8.5cm]{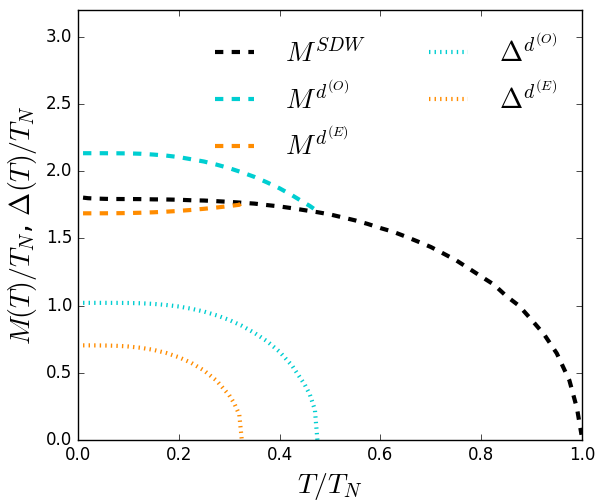}
\caption{Self-consistently calculated $M$ and $\Delta$ order parameters in the absence of superconductivity (black), when spin density waves coexist with a $d$-wave SC gap which is odd under translations of $\vQ$ (cyan), and when spin density waves coexist with a $d$-wave SC gap which is even under translations of $\vQ$ (orange).} 
\label{fig:deltaM}
\end{centering}
\end{figure}

The mean field order parameters $M$ and $\Delta$ can be self-consistently solved for from the Green's function method \cite{senchoudhury}$^,$\cite{kato}$^,$\cite{machida}$^,$\cite{machida2}$^,$\cite{kato2}. This can be obtained from the bare Matsubara Green's function which can be found from the Dyson equation:
\begin{equation}
    \hat{G}_\vk(\omega_n) = \big(i\omega_n \hat{I} - \hat{\mathcal{H}}_\vk \big)^{-1}
\end{equation}
\noindent where $\omega_n = 2\pi T (n + 1/2)$ is the Matsubara frequency. The relevant Green's functions for $\Delta$ are contained in the diagonal blocks, whereas the relevant Green's functions for $M$ are contained in the off-diagonal blocks. Calculating the relevant Green's functions from the Dyson equation and substituting them into the definitions of the mean-field order parameters $\Delta$ and $M$ yields two systems of equations for when the SC gap is odd or even under translations of $\vQ$. When the SC gap is odd under translations of $\vQ$, $1/V_{\text{SC}}$ and $1/V_{\text{SDW}}$ when $\Delta$ and $M$ are nonzero can be written as:

\begin{widetext}
\begin{equation}
\begin{split}
    \frac{1}{V_{\text{SC}}} &= T \sum_{\omega_n}^{E_c} \sum_{\vk} \frac{\mathcal{Y}_\vk^2}{D^{(O)}_\vk(\omega_n)} \big(\omega_n^2 + (\xi_\vk^-)^2 + (\xi_\vk^+)^2 + M^2 + \Delta_\vk^2 \big) \\
    \frac{1}{V_{\text{SDW}}} &= T \sum_{\omega_n}^{E_B} \sum_{\vk} \frac{1}{D^{(O)}_\vk(\omega_n)} \big(\omega_n^2 + (\xi_\vk^-)^2 - (\xi_\vk^+)^2 + M^2 + \Delta_\vk^2 \big) \\
    D^{(O)}_\vk &= \big(\omega_n^2 + (\xi_\vk^-)^2 + (\xi_\vk^+)^2 + \Delta_\vk^2 + M^2 \big)^2 - 4(\xi_\vk^+)^2\big((\xi_\vk^-)^2 + M^2 \big) \\
    &= \bigg(\omega_n^2 + \big(E^{(1;O)}_\vk \big)^2 \bigg) \bigg(\omega_n^2 + \big(E^{(2;O)}_\vk \big)^2 \bigg)\\
    \label{eq:MatsubaraOdd}
\end{split}
\end{equation}
\noindent and $1/V_{\text{SC}}$ and $1/V_{\text{SDW}}$ when the SC gap is even under translations of $\vQ$ when both $\Delta$ and $M$ are nonzero can be written as:
\begin{equation}
\begin{split}
    \frac{1}{V_{\text{SC}}} &= T \sum_{\omega_n}^{E_c} \sum_{\vk} \frac{\mathcal{Y}_\vk^2}{D^{(E)}_\vk(\omega_n)} \big(\omega_n^2 + (\xi_\vk^-)^2 + (\xi_\vk^+)^2 - M^2 + \Delta_\vk^2 \big) \\
    \frac{1}{V_{\text{SDW}}} &= T \sum_{\omega_n}^{E_B} \sum_{\vk} \frac{1}{D^{(E)}_\vk(\omega_n)} \big(\omega_n^2 + (\xi_\vk^-)^2 - (\xi_\vk^+)^2 + M^2 - \Delta_\vk^2 \big) \\
    D^{(E)}_\vk &= \big(\omega_n^2 + (\xi_\vk^-)^2 + (\xi_\vk^+)^2 + \Delta_\vk^2 + M^2 \big)^2 - 4(\xi_\vk^+)^2\big((\xi_\vk^-)^2 + M^2 \big) - 4M^2\Delta_\vk^2 \\
    &= \bigg(\omega_n^2 + \big(E^{(1;E)}_\vk \big)^2 \bigg) \bigg(\omega_n^2 + \big(E^{(2;E)}_\vk \big)^2 \bigg)
    \label{eq:MatsubaraEven}
\end{split}
\end{equation}
\end{widetext}

\noindent where $E_c = 2\pi T (n_c + 1/2)$ and $E_B = 2\pi T (n_B + 1/2)$ are the cutoff energies for the SC and SDW Matsubara sums respectively; $n_c = 30 T_N/T$ and $n_B = 175 T_N/T$ were used in this work. The natural choice of energy scale for these equations is $T_N$, since $T_c$ depends on the value of $M$. In order to self-consistently solve for the order parameters, $\Delta$ and $M$, the SC and SDW interaction potentials can be eliminated by subtracting Equation (\ref{eq:MatsubaraNormal}) from Equation (\ref{eq:MatsubaraOdd}) or Equation (\ref{eq:MatsubaraEven}). 
\begin{equation}
\begin{split}
    \frac{1}{V_{\text{SC}}} = T_c^0 \sum_{\omega_n}^{E_c} \frac{\mathcal{Y}_\vk^2 \big(\omega_n^2 + (\xi_\vk^-)^2 +(\xi_\vk^+)^2 \big)}{\big(\omega_n^2 + (\xi_\vk^-)^2 + (\xi_\vk^+)^2\big)^2 - 4(\xi_\vk^- \xi_\vk^+)^2} \\
    \frac{1}{V_{\text{SDW}}} = T_N \sum_{\omega_n}^{E_B} \frac{\big(\omega_n^2 + (\xi_\vk^-)^2 +(\xi_\vk^+)^2 \big)}{\big(\omega_n^2 + (\xi_\vk^-)^2 + (\xi_\vk^+)^2\big)^2 - 4(\xi_\vk^- \xi_\vk^+)^2} \\
    \label{eq:MatsubaraNormal}
\end{split}
\end{equation}
\noindent and $T_c^0$ is the superconducting transition temperature in the absence of spin density waves. In this work $T_c^0 = .35 T_N$ was used for both symmetry classes, but the actual superconducting transition temperatures were found to be $T^{(O)}_c = .47 T_N$ and $T^{(E)}_c = .32 T_N$ from self-consistency. The order parameters $\Delta$ and $M$ can be seen as a function of temperature for both the odd and even symmetry classes in FIG. ~\ref{fig:deltaM}.

\section{Two-Step Diagonalization}\label{sec:twoStepDiag}

A two-step process can be employed to simplify the calculation of the Bogoliubov transformation which diagonalizes the Hamiltonian when $M$ and $\Delta$ are simultaneously nonzero\cite{ismer}. The first step in this procedure is to diagonalize the Hamiltonian when $M \neq 0$ and $\Delta = 0$. It can be shown that the Bogoliubov transformation matrix in this case is $\hat{B}^{\text{SDW}}_\vk$, which can be used to define the states for the $E^{(\alpha)}_\vk$ and $E^{(\beta)}_\vk$ quasiparticle bands respectively as $\hat{\alpha}_{\vk,\sigma}^\dagger \ket{0}$ and $\hat{\beta}_{\vk,\sigma}^\dagger \ket{0}$:

\begin{widetext}
\begin{equation}
\begin{centering}
    \hat{\mathbf{\Psi}}^{\text{SDW}}_\vk = \hat{B}^{\text{SDW}}_\vk \hat{\mathbf{\Psi}}^{\text{n}}_\vk = 
    \begin{pmatrix}
    \hat{\alpha}_{\vk,\sigma} \\
    \hat{\alpha}_{-\vk,-\sigma}^\dagger \\
    \hat{\beta}_{\vk,\sigma} \\
    \hat{\beta}_{-\vk,-\sigma}^\dagger \\
    \end{pmatrix} = \begin{pmatrix}
    u_\vk & 0 & v_\vk & 0 \\
    0 & u_\vk & 0 & -v_\vk \\
    -v_\vk & 0 & u_\vk & 0 \\
    0 & v_\vk & 0 & u_\vk \\
    \end{pmatrix} \begin{pmatrix}
    \hat{a}_{\vk,\sigma} \\
    \hat{a}_{-\vk,-\sigma}^\dagger \\
    \hat{a}_{\vk + \vQ,\sigma} \\
    \hat{a}_{-\vk-\vQ,-\sigma}^\dagger
    \end{pmatrix}
\end{centering}
\end{equation}

\end{widetext}
\noindent where $u_\vk = \sqrt{\frac{1}{2}(1+\frac{\xi_\vk^-}{\zeta_\vk})}$, $v_\vk = \sqrt{\frac{1}{2}(1-\frac{\xi_\vk^-}{\zeta_\vk})}$, and $\zeta_\vk = \sqrt{(\xi_\vk^-)^2 + M^2}$. The transformation matrix $\hat{B}^{\text{SDW}}_\vk$ is used to calculate the coherence matrix in Equation (\ref{eq:coherence}) when $T > T_c$ in the regions of $\vk$-space in which the FS reconstruction occurs. However, this transformation matrix can also be used to rewrite the Hamiltonian, $\hat{\mathcal{H}}_\vk$, when both $M$ and $\Delta$ are nonzero in the basis of the SDW Nambu vector, $\hat{\mathbf{\Psi}}^{\text{SDW}}_\vk$, as $\hat{\mathcal{H}}'_\vk$:

\begin{equation}
    H = \frac{1}{2} \sum_{\vk}\big(\hat{\mathbf{\Psi}}_{\vk}^{\text{SDW}}\big)^{\dag} \hat{\mathcal{H}}'_{\vk} \hat{\mathbf{\Psi}}_{\vk}^{\text{SDW}}
\end{equation}

\noindent Performing this change of basis on the coexistence Hamiltonian without loss of generality results in a Hamiltonian with intraband coupling terms, which couple the $E^{(\alpha,\beta)}_\vk$ bands with the $-E^{(\alpha,\beta)}_\vk$ bands, and interband coupling terms, which couple the $E^{(\alpha,\beta)}_\vk$ bands with the $-E^{(\beta,\alpha)}_\vk$ bands.

\begin{widetext}
\begin{equation}
\begin{split}
    \hat{\mathcal{H}}'_{\vk} &= \hat{B}_{\vk}^{\text{SDW}} \hat{\mathcal{H}}_{\vk} \big(\hat{B}_{\vk}^{\text{SDW}}\big)^\dag \\
    &= \begin{pmatrix}
        E^{(\alpha)}_\vk & \Delta_\vk\abs{u_\vk}^2 - \Delta_{\vk+\vQ}\abs{v_\vk}^2 & 0 & 2\Delta_\vk^+ u_\vk v_\vk \\
        \Delta_\vk \abs{u_\vk}^2 - \Delta_{\vk+\vQ} \abs{v_\vk}^2 & -E^{(\alpha)}_\vk & -2\Delta_\vk^+ u_\vk v_\vk & 0\\
        0 & -2\Delta_\vk^+ u_\vk v_\vk & E^{(\beta)}_\vk & -\Delta_\vk \abs{v_\vk}^2 + \Delta_{\vk+\vQ} \abs{u_\vk}^2\\
        2\Delta_\vk^+ u_\vk v_\vk & 0 & -\Delta_\vk \abs{v_\vk}^2 + \Delta_{\vk + \vQ} \abs{u_\vk}^2 & -E^{(\beta)}_\vk\end{pmatrix}
\end{split}
\end{equation}
\end{widetext}

\noindent Analytically diagonalizing the general equation for the Hamiltonian in the $\hat{\mathbf{\Psi}}^{\text{SDW}}_\vk$ basis is a difficult task, however it can be simplified in the cases of SC gaps with even and odd symmetries under translations of $\vQ$. The simplest case to consider is the odd case, where $\Delta_{\vk+\vQ} = -\Delta_\vk$ which simplifies the intraband coupling terms to $\pm \Delta_\vk \big(\abs{u_\vk}^2 + \abs{v_\vk}^2\big) = \pm \Delta_\vk$. This also reduces the interband coupling terms to $0$, decoupling the $E^{(\alpha)}_\vk$ and $E^{(\beta)}_\vk$ SDW bands entirely. Therefore, the Hamiltonian in the $\hat{\mathbf{\Psi}}^{\text{SDW}}_\vk$ basis when the SC gap is odd under translations of $\vQ$ reduces to:

\begin{equation}
    \hat{\mathcal{H}}'_{\vk} = \begin{pmatrix}
        E^{(\alpha)}_\vk & \Delta_\vk & 0 & 0 \\
        \Delta_\vk & -E^{(\alpha)}_\vk & 0 & 0\\
        0 & 0 & E^{(\beta)}_\vk & -\Delta_\vk\\
        0 & 0 & -\Delta_\vk & -E^{(\beta)}_\vk\end{pmatrix}
\end{equation}

\noindent which can be diagonalized by two separate Bogoliubov transformations with dispersion relations $\epsilon^{(1;O)}_\vk = \sqrt{\big(E^{(\alpha)}_\vk \big)^2 + \Delta_\vk^2}$ and $\epsilon^{(2;O)}_\vk = \sqrt{\big(E^{(\beta)}_\vk \big)^2 + \Delta_\vk^2}$, which can be shown\cite{senchoudhury} to be equivalent to $E^{(1;O)}_\vk$ and $E^{(2;O)}_\vk$. These Bogoliubov transformations can also be used to define the states for the $E^{(1)}_\vk$ and $E^{(2)}_\vk$ as $\hat{c}_{1,\vk,\sigma}^\dagger \ket{0}$ and $\hat{c}_{2,\vk,\sigma}^\dagger \ket{0}$ respectfully by performing the transformation on $\hat{\mathbf{\Psi}}^{\text{SDW}}_\vk$:

\begin{widetext}
\begin{equation}
    \hat{\mathbf{\Psi}}^{(O)}_\vk = \hat{\mathcal{B}}_\vk^{(O)} \hat{\mathbf{\Psi}}^{\text{SDW}}_\vk = \begin{pmatrix}
        \hat{c}_{1,\vk,\sigma} \\
        \hat{c}_{1,-\vk,-\sigma}^\dagger \\
        \hat{c}_{2,\vk,\sigma} \\
        \hat{c}_{2,-\vk,-\sigma}^\dagger \\
    \end{pmatrix} = \begin{pmatrix}
        u^{(1;O)}_\vk & v^{(1;O)}_\vk & 0 & 0 \\
        -v^{(1;O)}_\vk & u^{(1;O)}_\vk & 0 & 0 \\
        0 & 0 & u^{(2;O)}_\vk & -v^{(2;O)}_\vk \\
        0 & 0 & v^{(2;O)}_\vk & u^{(2;O)}_\vk \\
    \end{pmatrix} \begin{pmatrix}
    \hat{\alpha}_{\vk,\sigma} \\
    \hat{\alpha}_{-\vk,-\sigma}^\dagger \\
    \hat{\beta}_{\vk,\sigma} \\
    \hat{\beta}_{-\vk,-\sigma}^\dagger \\
    \end{pmatrix}
\end{equation}

\end{widetext}
\noindent where $u^{(1,2;O)}_\vk = \sqrt{\frac{1}{2}\big(1 + \frac{E^{(\alpha,\beta)}_\vk}{E^{(1,2;O)}_\vk}\big)}$ and $v^{(1,2;O)}_\vk = \sqrt{\frac{1}{2}\big(1 - \frac{E^{(\alpha,\beta)}_\vk}{E^{(1,2;O)}_\vk}\big)}$. $\hat{B}^{(O)}_\vk = \hat{\mathcal{B}}^{(O)}_\vk \hat{B}^{\text{SDW}}_\vk$ is used to calculate the coherence factor in Equation (\ref{eq:coherence}) when $\vk$ is in the region where the FS becomes reconstructed, $\Delta \neq 0$, and $\Delta_\vk$ is odd under translations of $\vQ$. The $\hat{B}_\vk^{(O)}$ transformation matrix calculated here is consistent with previous calculations in literature for the cuprates\cite{das} if the order of the Nambu vector elements are properly accounted for. When the SC gap is even under translations of $\vQ$, $\hat{\mathcal{H}}'_\vk$ can't be simplified generally beyond setting $\Delta_{\vk+\vQ} = \Delta_\vk$, and $\hat{\mathcal{H}}'_\vk$ needs to be diagonalized numerically in order to calculate $\hat{\mathcal{B}}^{(E)}_\vk$.

%%%%%%%%%%%%%%%%%%%%%%%%%%%%%%%%%%%%%%%%%%%%%%%%%%%%%%%%%%%%%%%%%%%

\bibliography{ref}

\bibliographystyle{apsrev.bst}

\end{document}